\documentclass[%
reprint,
superscriptaddress,
amsmath,amssymb,
aps,
prb,
]{revtex4-2}
\usepackage{graphicx}
\usepackage{dcolumn}
\usepackage{bm}
\usepackage{hyperref}

\usepackage{xcolor}

\begin{document}
	\title{Topological superconductivity in quasicrystals}
	\author{Rasoul Ghadimi}
	\affiliation{Department of Applied Physics, Tokyo University of Science, Tokyo 125-8585, Japan}
	\author{Takanori Sugimoto}
	\affiliation{Department of Applied Physics, Tokyo University of Science, Tokyo 125-8585, Japan}
        \affiliation{
          Advanced Science Research Center, Japan Atomic Energy Agency, Tokai, Ibaraki 319-1195, Japan}
        \author{K. Tanaka}
        \affiliation{Department of Physics and Engineering Physics, and Centre for Quantum Topology and Its Applications (quanTA), University of Saskatchewan, 116 Science Place, Saskatoon, Saskatchewan, Canada S7N 5E2}
	\author{Takami Tohyama}
	\affiliation{Department of Applied Physics, Tokyo University of Science, Tokyo 125-8585, Japan}
	\date{Published 26 October 2021}

	\begin{abstract}
          We propose realization of non-Abelian topological superconductivity in two-dimensional quasicrystals by the same mechanism as in crystalline counterparts.
	  Specifically, we study a two-dimensional electron gas in Penrose and Ammann-Beenker quasicrystals with Rashba spin-orbit coupling, perpendicular Zeeman magnetic field, and conventional $s$-wave superconductivity. 
          We find that topological superconductivity with broken time-reversal symmetry is realized in both Penrose and Ammann-Beenker quasicrystals at low filling, where the Bott index is unity. 
          The topological nature of this phase is confirmed by the existence of a zero-energy surface bound state and the chiral propagation of a wave packet projected onto the midgap bound state along the surfaces.
          Furthermore, we confirm the existence of a single Majorana zero mode each in a vortex at the center of the system and along the surfaces, signifying the non-Abelian character of the system when the Bott index is unity.
	\end{abstract}
	\maketitle

	\begin{figure}
	  \centering
	  \includegraphics[width=\linewidth]{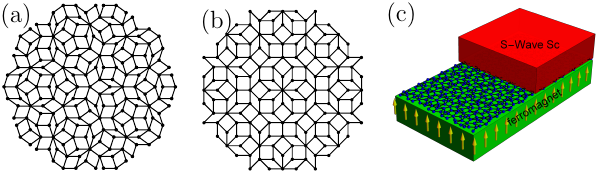}
	  \caption{(a) Penrose and (b) Ammann-Beenker quasicrystals studied in this work. 
            (c) A schematic setup for realizing a topological quasicrystal superconductor in heterostructure. 
          }
	  \label{fig:crystall}
	\end{figure}

        \section{Introduction}
        Since the first discovery of three-dimensional topological insulators about a decade ago \cite{RevModPhys.82.3045,doi:10.1146/annurev-conmatphys-062910-140432}, a wide variety of topological materials have been discovered theoretically as well as experimentally \cite{RevModPhys.83.1057,RepProgPhys.75.076501,annurev-conmatphys-031214-014501,0034-4885-80-7-076501}. Classification of topological materials \cite{PhysRevB.78.195125} is not only limited to crystalline systems, but also has been extended to include disordered \cite{Loring_2010}, amorphous \cite{Mitchell2018}, and quasicrystal materials. 
        Quasicrystals \cite{PhysRevB.34.596,PhysRevB.34.617}, which present phases of matter with long-range structural order without periodicity \cite{PhysRevLett.53.1951,PhysRevLett.53.2477}, have brought about new research into topology in condensed matter systems \cite{PhysRevLett.109.106402,PhysRevB.88.125118}.
        Topological properties of quasicrystals \cite{PhysRevLett.110.076403,PhysRevLett.119.215304} have been investigated in connection with the quantum Hall effect \cite{PhysRevLett.111.226401,PhysRevB.91.085125,PhysRevB.98.165427}, the quantum spin Hall effect \cite{PhysRevLett.121.126401,PhysRevB.98.125130,PhysRevB.100.085119}, higher-order topological phases \cite{chen2019higherorder,PhysRevLett.123.196401}, and superconductivity \cite{PhysRevLett.116.257002,PhysRevB.95.024509,Kamiya2018,PhysRevResearch.1.022002,PhysRevB.100.014510}.
        Moreover, recent technical advances for experimental realization of two-dimensional (2D) quasicrystals, either in optical lattices \cite{PhysRevLett.79.3363,PhysRevX.6.011016} or by means of crystal growth technologies \cite{PhysRevLett.92.135507,PhysRevB.72.045428,PhysRevB.77.073409,PhysRevB.78.075407,Forster2013,Collins2017}, have enabled studies of novel quantum phenomena in actual quasicrystals.

	
        One of the most significant properties of topological materials is the existence of Majorana fermions in topological superconductors \cite{RepProgPhys.75.076501}. The capability of creating and manipulating Majorana fermions in a solid device may well open the door to realizing stable and scalable quantum computation that is topologically protected \cite{0034-4885-80-7-076501}.
        In a one-dimensional system presenting topological superconductivity (TSC) as in the Kitaev model \cite{Kitaev_2001}, zero-energy Majorana fermions appear at the two ends of the system 
        \cite{Mourik2012,NadjPerge2014}.
        In a 2D topological superconductor, a Majorana zero mode can appear not only along a surface \cite{natcomm8.2040.2017}, but also in the vortex core \cite{PhysRevLett.116.257003}. 2D TSC with broken time-reversal symmetry has been proposed to be realized in an ultracold Fermi gas \cite{PhysRevLett.103.020401,PhysRevB.82.134521} and heterostructure made of conventional materials \cite{PhysRevLett.104.040502,PhysRevB.81.125318,RepProgPhys.75.076501}, and has been achieved in a Pb/Co island on Si(111) \cite{natcomm8.2040.2017}.
        Necessary ingredients are 2D $s$-wave superconductivity \cite{PhysRevB.89.184514,scheurer2015topological,GARIGLIO2015189,saito2017highly}, Rashba spin-orbit coupling (RSOC) \cite{Rashba60,BychkovRashba84,gor2001superconducting}, and perpendicular Zeeman magnetic field (PZMF). RSOC 
        can be enhanced or induced by proximity effects in heterostructures \cite{PhysRevLett.115.147003,Nature.529.185,PhysRevB.95.064509,Soumyanarayanan2016,Han2018}.

	In our previous work \cite{doi:10.7566/JPSJ.86.114707}, we have studied the topological phase diagram of the Fibonacci-Kitaev model as an example of the simplest one-dimensional quasicrystalline topological superconductors. We have found that quasicrystal structure has a profound effect on the topological phase diagram, making it fractal. One might now ask, what will happen if the spatial dimension increases from one to two? Can TSC be stable even in 2D quasicrystals? In order to answer these questions, in this work we apply the method of realizing 2D TSC with broken time-reversal symmetry \cite{PhysRevLett.103.020401,PhysRevB.82.134521} 
        to quasicrystals. Specifically, we study Penrose \cite{Penrose} and Ammann-Beenker (AB) \cite{grunbaum1987tilings,Ammann1992,beenker1982algebraic} quasicrystals [see Figs.~\ref{fig:crystall}(a) and \ref{fig:crystall}(b)] at low filling with RSOC, PZMF, and $s$-wave superconducting pairing. 
	We find that irrespective of the aperiodicity of a quasicrystal, TSC is realized as in a square lattice with translational invariance. This finding is obtained by calculating the Bott index as a topological invariant in the system \cite{PhysRevLett.121.126401,PhysRevB.98.125130,Loring_2010} and 
        confirming the existence of a Majorana zero mode in a vortex and along the surfaces in the topological phase where the Bott index is unity.

        The paper is organized as follows. The model is described in Sec.~\ref{sec:model}, results are presented and discussed in Sec.~\ref{sec:results}, and the work is summarized in Sec.~\ref{sec:conclusion}. We describe our method of producing AB approximants in the Appendix. 

	\section{Model} \label{sec:model}
        We focus on 2D Penrose and AB 
quasicrystals as illustrated in Figs.~\ref{fig:crystall}(a) and \ref{fig:crystall}(b). Our results can be generalized readily for other types of 2D quasicrystals. 
        We generalize the tight-binding model
        \cite{PhysRevLett.103.020401,PhysRevB.82.134521} 
        for a quasicrystal:
	\begin{equation}
	\label{Eq:BdG}
	{\cal H} = \frac{1}{2}\sum_{ij\sigma\sigma^\prime}\left(\begin{matrix}
	c_{i\sigma}^\dagger & c_{i\sigma} 
	\end{matrix}\right)H
        \left(\begin{matrix}
	c_{j\sigma^\prime} \\ c_{j\sigma^\prime}^\dagger 
	\end{matrix}\right),        
        \qquad H=\left(\begin{matrix}
	\mathbf h& \mathbf\Delta \\	\mathbf\Delta^\dagger&-\mathbf h^* 
	\end{matrix}\right),
	\end{equation}
	where $c_{i\sigma}$ is the annihilation operator for the electron at site $i$ with spin $\sigma$, and the normal-state Hamiltonian is
	\begin{eqnarray}
	[\mathbf h]_{i\alpha,j\beta}=\left[ (t_{ij}-\mu \delta_{ij}) \sigma_0+h_z\delta_{ij} \sigma_3 + \imath V_{ij} \vec{e}_z\cdot \vec{\sigma}\times \hat{R}_{ij} \right]_{\alpha \beta},
	\end{eqnarray}
	with the Pauli matrices $\vec \sigma=(\sigma_1,\sigma_2,\sigma_3)$ acting in spin space, 
        $\sigma_0={\bf 1}_{2}$ the $2\times 2$ identity matrix, spin indices $\alpha,\beta$, and $\imath=\sqrt{-1}$.
	We consider the vertex model, 
where the sites $\{i\}$ are defined on vertices in the quasicrystal and $\hat{R}_{ij}$ is a unit vector connecting 
        sites $i$ and $j$.
	We consider hopping along nearest-neighbor links only, $t_{ij}=t_{\langle ij\rangle}\equiv -t$, and $V_{ij}=V_{\langle ij\rangle}\equiv V$ is the coupling constant of RSOC, where $\langle \rangle$ indicates nearest-neighbor links.
	PZMF and the chemical potential are denoted as $h_z$ and $\mu$, respectively.
        The off-diagonal elements are given by
	\begin{equation}\label{Eq:pairing}
	[\mathbf \Delta]_{i\alpha,j\beta}=\left[\delta_{ij} \Delta\imath \sigma_{2}\right]_{\alpha \beta},
	\end{equation}
	where 
        $\Delta$ is the $s$-wave superconducting order parameter.
	A possible setup of the system is illustrated in Fig.~\ref{fig:crystall}(c), where PZMF and $s$-wave superconductivity are induced by proximity to a ferromagnetic insulator and a conventional superconductor, respectively.
        RSOC can be enhanced or induced by the ferromagnetic insulator \cite{PhysRevLett.115.226601} or the superconductor \cite{grapheneWS2,grapheneMoS2}.
        To explore the properties of such a system, we numerically diagonalize the BdG Hamiltonian in Eq.~(\ref{Eq:BdG}) to find the quasiparticle energy spectrum and wave functions \cite{deGennes}:
	\begin{equation}
		H \left|\psi_{\lambda}\right>=\epsilon_{\lambda}\left|\psi_{\lambda}\right>.
                \label{Eq:BdGeq}
	\end{equation}
	 
        The topological phases in a square lattice with translational symmetry have been classified in Ref.~\cite{PhysRevB.82.134521} according to the first Chern number or the Thouless-Kohmoto-Nightingale-Nijs (TKNN) number 
        \cite{TKNN1982}, $\nu\in\mathbb{Z}$ \cite{PhysRevB.78.195125}, where the system is in the trivial, Abelian, and non-Abelian phase when $\nu=0$, $\nu=-2$, and $\nu=\pm 1$, respectively. For the chemical potential $\mu\le -2t$ and for large enough PZMF, the system can have a single noninteracting Fermi surface, and the non-Abelian phase with $\nu=1$ is realized when $\Delta^2<h_z^2-(W+\mu)^2$, where $\Delta$ is taken to be real and $W=4t$ is half of the bandwidth in the absence of RSOC and PZMF in the normal state.
 This phase 
        hosts a zero-energy Majorana fermion as a single edge mode per surface or bound state in a vortex \cite{PhysRevB.77.220501,PhysRevB.82.134521,PhysRevB.94.064515}. 
        
        In the following, we set 
        \begin{equation}
          V=0.5t,\quad h_z=0.5t,\quad \Delta=0.2t,
          \label{parameters}
        \end{equation}
        and probe topological phase transitions by varying the chemical potential in the low-filling limit.
    
        The Bott index \cite{Loring_2010,doi:10.1063/1.5083051} is one of the topological invariants that are equivalent to the first Chern number, previously used \cite{PhysRevLett.121.126401,PhysRevB.98.125130,PhysRevB.100.085119,PhysRevLett.116.257002,PhysRevX.6.011016} to explore nontrivial states of a quasicrystal. In order to calculate the Bott index, we first obtain the quasiparticle excitation states. 
        Exploiting the particle-hole symmetry of Eq.~(\ref{Eq:BdGeq}), we define the occupation projector onto the quasiparticle states with negative energy,
	\begin{equation}\label{Eq:Projector}
	P=\sum_{\epsilon_{\lambda}<0} \left|\psi_{\lambda}\right> \left<\psi_{\lambda}\right|.
	\end{equation}
        In terms of this projector and $Q=I-P$, with $I$ the identity operator, 
        we can define the projected position operators,
	\begin{equation}\label{ProjectedPositionOperator}
	U_X=P e^{\imath 2\pi X}P+Q,\qquad 		U_Y=P e^{\imath 2\pi Y}P+Q,
	\end{equation} 
	where 
	\begin{equation}\label{Eq:PositionOperator}
	X=\text{Diag}[x_1,x_1,\dots,x_N,x_N,x_1,x_1,\dots,x_N,x_N]\,.
	\end{equation}
	Here $N$ is the total number of vertices, $x_i$ is the $x$ coordinate of the $i$th vertex rescaled to $\left[0,1\right)$, and similarly for $Y$. 
        Namely, each vertex (lattice site) of a 2D system is mapped onto the surface of a torus.
        The Bott index is defined by
	\begin{equation}\label{Eq:BottIndex}
	B=\frac{1}{2\pi} \text{Im} \big(\text{Tr}\big[\log(U_YU_XU_Y^\dagger U_X^\dagger)\big]\big),
	\end{equation}
	which is quantized to be a nonzero integer (zero) in a
        topologically nontrivial (trivial) phase. We use the periodic boundary condition (PBC) for large enough system size to calculate the Bott index. In nontrivial topological states the periodic and open boundary conditions (OBC) result in a gapful and gapless energy spectrum, respectively. The latter is a direct consequence of the bulk-boundary correspondence \cite{RevModPhys.88.035005}. To apply PBC to AB 
quasicrystal supercells, we first identify a large square portion of the quasicrystal that has similar edges, and then apply PBC to each pair of parallel edges. Our method of generating AB approximants is described in detail in the Appendix. For Penrose quasicrystals we use the multigrid method \cite{deBruijn,PhysRevB.43.8879}. Details of our method of generating Penrose approximants are explained in the Appendix of Ref.~\cite{ghadimi2020}.
	
	\begin{figure}
	  \centering
          \includegraphics[width=0.9\columnwidth]{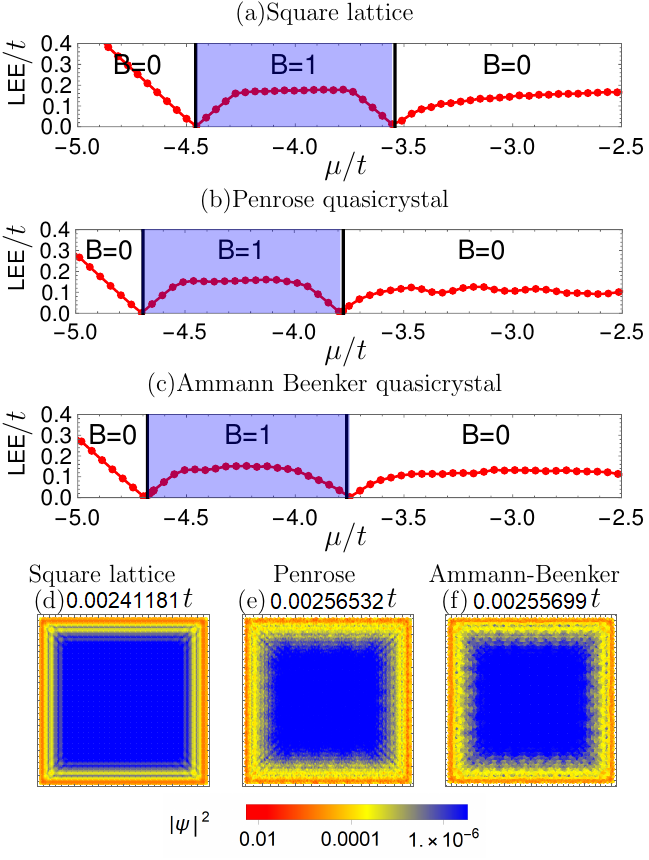}
	  \caption[Gap and Bott index]{Lowest excitation energy (LEE) is plotted as a function of the chemical potential for (a) a square lattice ($2916$ vertices) and (b) Penrose ($3571$ vertices) and (c) Ammann-Beenker ($2869$ vertices) quasicrystals with PBC. 
            The probability distribution of the lowest-energy excitation in a (d) square lattice ($18225$ vertices) and (e) Penrose ($18643$ vertices) and (f) Ammann-Beenker ($18029$ vertices) quasicrystals with OBC 
            for $\mu=-4.25t$, in a logarithmic scale.
            The darker (red) color implies higher probability. The energy of each state is shown above each plot.}
	  \label{fig:evbs}
	\end{figure}
	
	\begin{figure}
	  \includegraphics[width=0.9\columnwidth]{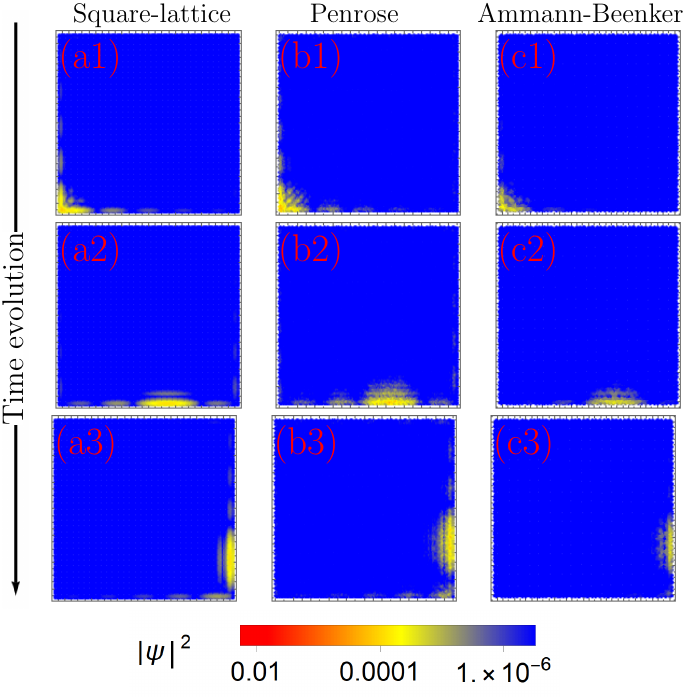}
	  \caption[Time evolution of state]{Time evolution of the chiral propagation of a surface bound state 
            is illustrated for the same systems as in Fig.~\ref{fig:evbs}(d)--\ref{fig:evbs}(f) for (a1)--(a3) a square lattice and (b1)--(b3) Penrose and (c1)--(c3) Ammann-Beenker quasicrystals 
for $\mu=-4.25t$. 
          }
	  \label{fig:Pro}
	\end{figure}

        \section{Results} \label{sec:results}
        \subsection{Topological phase transitions} \label{subsec:PT}
	In Fig.~\ref{fig:evbs}(a)--\ref{fig:evbs}(c) we present the Bott index and the lowest quasiparticle excitation energy as a function of the chemical potential $\mu$ for a $54\times 54$ square lattice and 
        Penrose ($3571$ vertices) and AB 
($2869$ vertices) quasicrystals.
	It can be seen that irrespective of the crystal structure, the energy gap closes twice as $\mu$ is increased in the region shown, where the Bott index $B$ changes first from zero to unity and then back to zero. We find that the range of $\mu$ for which $B=1$ is given by
        \begin{equation}
          -W-\sqrt{h_z^2-\Delta^2}<\mu<-W+\sqrt{h_z^2-\Delta^2}\,,
          \label{murange}
        \end{equation}
        where $-W$ is the lower band edge in the absence of RSOC and PZMF in the normal state, regardless of the crystal structure. With $W=4t$, this is precisely the condition for the non-Abelian phase with the TKNN number $\nu=1$ in a square lattice \cite{PhysRevB.82.134521}. The two critical values of $\mu$ above are indicated by two vertical lines for each system in Fig.~\ref{fig:evbs}(a-c).
        We have confirmed these phase transitions for different combinations of parameter values ($V$, $h_z$, $\Delta$) and system size.
	
        The bulk-boundary correspondence implies the existence of a gapless bound state per surface in the topological phase with $B=1$. This is illustrated in Fig.~\ref{fig:evbs}(d-f), where the probability distribution is plotted for the lowest-energy state in a $135\times 135$ square lattice and 
        Penrose ($18643$ vertices) and AB 
($18029$ vertices) quasicrystals for $\mu=-4.25t$. The energy ($\sim 10^{-3}t$) for each state is shown above each plot.
        Clearly these states are strongly localized along the surfaces, and the energy of these states approaches zero as the system size increases.
        In the thermodynamic limit, these midgap surface bound states form a continuous excitation spectrum.
        In contrast, in the trivial phase there is no such surface bound state and the wave function distribution depends drastically on $\mu$ 
        and the shape of the system.
		
	\begin{figure*}[t!]
	  \includegraphics[width=0.85\linewidth]{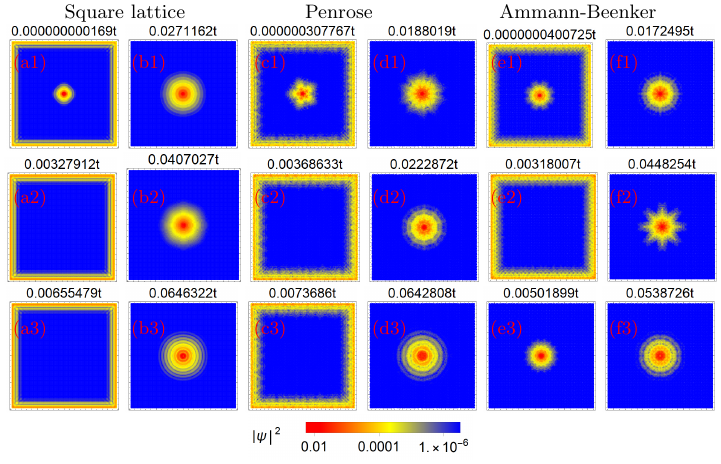}
	  \caption[Vortex]{The probability distribution of the three lowest-energy excitations is plotted in a logarithmic scale for a square lattice in the (a1)--(a3) topologically nontrivial ($\mu=-4.25t$) and (b1)--(b3) trivial ($\mu=-3t$) phase, Penrose quasicrystal in the (c1)--(c3) topologically nontrivial ($\mu=-4.25t$) and (d1)--(d3) trivial ($\mu=-3.5t$) phase, and Ammann-Beenker quasicrystal in the (e1)--(e3) topologically nontrivial ($\mu=-4.25t$) and (f1)--(f3) trivial ($\mu=-3.5t$) phase.
The numbers of vertices are $38025$, $37351$ and $38413$, respectively.
            The excitation energy of each state is shown above each plot.}
	  \label{fig:Majorana}
        \end{figure*}

        \subsection{Chiral propagation} \label{subsec:chiral}
        Because of the chiral nature of edge modes in a square lattice \cite{PhysRevB.77.220501,PhysRevB.82.134521}, we anticipate the unidirectional propagation of the midgap surface bound states. To see this, we project an initial state $\left|\psi_0\right>$ localized around an edge site onto the lowest-energy midgap state.
        We then allow the system to evolve with time by applying the time evolution operator $\exp(-\imath  T H)$ at time $T$. If the system supports chiral edge modes, the initial state would propagate along the boundary \cite{PhysRevB.91.085125,agarwala2018fractalized,PhysRevX.6.011016}.
	In Fig.~\ref{fig:Pro} the time-lapse propagation of the initial state
        is presented for the first few time steps, $\Delta T=50/t$. We can see that despite the aperiodicity, the wave packet propagates along the surface boundary in both 
        quasicrystals as in a square lattice.
        On the contrary, in trivial phase an initial wave packet quickly disperses into the bulk of the system.

	\subsection{Majorana zero modes} \label{subsec:majorana}
        It is possible to directly confirm the existence of Majorana zero modes by introducing a vortex in the system. For this purpose, we include a vortex as a local phase winding in the order parameter 
        in the middle of the crystal. We set the pairing amplitude to zero at the vortex center to avoid ambiguity in the pairing phase, while assuming no radial dependence in the amplitude or phase. 
	Introducing a vortex induces
        the Caroli-de Gennes-Matricon (CdGM) bound states \cite{CAROLI1964307,deGennes} localized in the vortex core.
        In the $B=1$ phase, 
        we additionally find a zero-energy Majorana bound state in the vortex core, which is clearly distinct from the CdGM states, as its energy is approximately zero regardless of 
        $\mu$ and approaches zero as the system size increases. In contrast, the CdGM energy levels in either trivial or nontrivial phase are strongly dependent on $\mu$.
        Moreover, while the CdGM excitation energy also depends on the local environment of the vortex center in quasicrystals, the Majorana bound-state energy does not.

        With a vortex at the center of the system with OBC, 
        the BdG equations (\ref{Eq:BdGeq}) yield two zero-energy solutions, numerically with energy $\pm \epsilon$ where $\epsilon \ll t$. 
	In Fig.~\ref{fig:Majorana} we plot the probability distribution of the three lowest-positive-energy states for two values of $\mu$ each for a square lattice ($38025$ vertices) and Penrose ($37351$ vertices) and AB 
($38413$ vertices) quasicrystals, such that $B=1$ for one value of $\mu$ ($\mu=-4.25t$ for all systems) and $B=0$ for the other ($\mu=-3t$ for the square lattice and $\mu=-3.5t$ for the quasicrystals). The energy of each state is shown above each plot. The highest-symmetry (maximum) coordination number is five (seven) and eight (eight), respectively, in the Penrose and AB 
quasicrystal. The vortex is placed at a highest-symmetry vertex. 
        The zero-energy state in all three systems shown in Fig.~\ref{fig:Majorana}(a1,c1,e1) has half of its probability distributed along the surfaces and the other half concentrated around the vortex center. This is also the case for the other zero-energy state (numerically with slightly negative energy) in each system. Furthermore, we have confirmed (not shown) that each 
zero-energy state has equal probabilities being an electron and a hole.
        Thus, analogously to the non-Abelian phase in the square lattice, a Majorana zero mode exists per vortex or surface in both kinds of quasicrystals in the topological phase with $B=1$.
        Interestingly, the third excitation 
in Fig.~\ref{fig:Majorana}(e3) is a CdGM state, while it can be a surface state 
depending on $\mu$ and the position of the vortex center.
We find that the highest coordination number results in the lowest energy of a given CdGM state. It is the combination of the highest coordination number and highest (eightfold) symmetry of the vortex center that lowers the energy of the CdGM excitation in the AB 
quasicrystal. 

	\begin{figure}
	  \includegraphics[width=0.95\columnwidth]{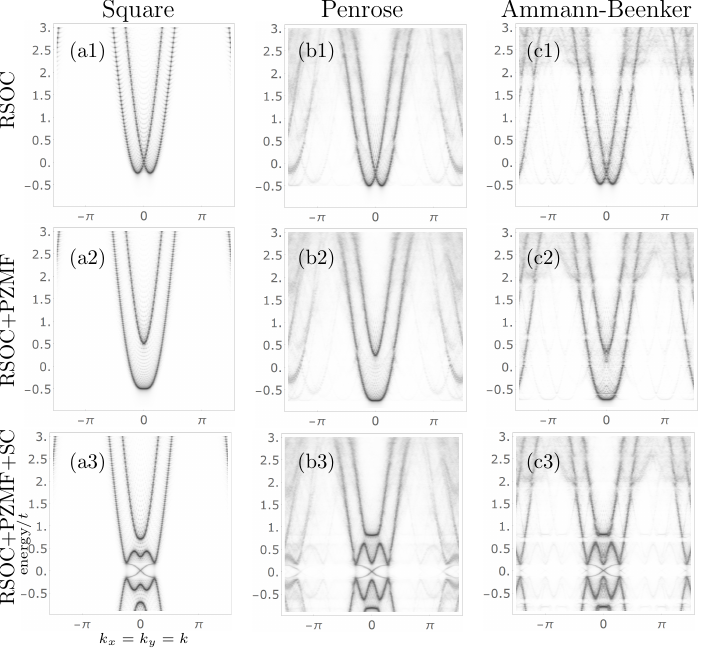}
	  \caption[Dispersion]{The probability distribution of single-particle states along $k_x=k_y$ in momentum space; in the normal state with RSOC [(a1), (b1), (c1)] and RSOC+PZMF [(a2), (b2), (c2)], and in the superconducting state [(a3), (b3), (c3)] for a square lattice [(a1)--(a3)] and Penrose [(b1)--(b3)] and Ammann-Beenker [(c1)--(c3)] quasicrystals with OBC with 3249, 3274, and 3105 vertices, respectively. The chemical potential $\mu=-4t$ is used for both normal and superconducting states. The opacity is given in a logarithmic scale.}
	  \label{fig:Dispersion}
        \end{figure}

        \subsection{Dispersion} \label{subsec:dispersion}

It is rather remarkable that the topological phase transitions as seen in Fig.~\ref{fig:evbs}(a-c) are predicted by Eq.~(\ref{murange}), which corresponds to closing of the bulk spectral gap at one of the high-symmetry points in the Brillouin zone in a square lattice, with numerically obtained $W$ in both quasicrystals. Due to the lack of periodicity, there is no Brillouin zone for quasicrystals and the entire momentum space would be filled with Bragg peaks in the limit of an infinite quasiperiodic lattice. Moreover, quasicrystals are known to have a pseudogap in their energy spectrum \cite{PhysRevB.40.942} (see also Refs.~\cite{Collins2017,PhysRevB.100.014510} and references therein). In addition, both Penrose and AB quasicrystals have families of strictly localized states at zero energy owing to local topology and quasiperiodicity \cite{PhysRevLett.56.2740,PhysRevB.51.15827,Grimm_Schreiber2002}, and in the former there is an energy gap above and below the highly degenerate ($\sim 10$\% of the total number of states) zero-energy level \cite{PhysRevLett.56.2740,PhysRevB.38.1621}. On the other hand, most of the kinetic-energy eigenstates in both quasicrystals are of the extended nature \cite{Grimm_Schreiber2002,PhysRevB.100.014510,PhysRevB.95.024509}. 

To better understand the occurrence of TSC governed by Eq.~(\ref{murange}), we now examine the band structure of quasicrystals. We first note that as we consider only nearest-neighbor links in the vertex model, all the links connecting vertices are of the same length in each of the Penrose and AB quasicrystals, and both systems are bipartite. As a result, the eigenspectrum of the hopping matrix is symmetric about zero energy in both quasicrystals \cite{PhysRevLett.56.2740,Grimm_Schreiber2002,PhysRevB.102.064210}. 

In Fig.~\ref{fig:Dispersion} we show the dispersion along $k_x=k_y$ for a square lattice [panels (a1)--(a3)] and Penrose [panels (b1)--(b3)] and AB [panels (c1)--(c3)] quasicrystals, with 3249, 3274, and 3105 vertices, respectively, with OBC. The dispersion has been obtained by calculating the spectral function \cite{PhysRevB.100.014510}, that is, the probability distribution of single-particle states in momentum space. In Fig.~\ref{fig:Dispersion}, 
the probability distribution of single-particle states is plotted as a function of $k_x=k_y$ and energy in the normal state with RSOC [panels (a1), (b1), (c1)] and RSOC+PZMF [panels (a2), (b2), (c2)], and in the superconducting state [panels (a3), (b3), (c3)], with the parameter values in Eq.~(\ref{parameters}). In all three systems, $\mu=-4t$ for both normal and superconducting states, and the smoothing width of $0.005t$ has been used in the spectral function. The probability is represented in a logarithmic scale by the opacity such that the darker a given point is, the higher the probability. The single-particle states in both quasicrystals extend over the entire momentum range shown. However, the probabilities are the highest in the energy range close to the bottom of the band in the normal state within the first Brillouin zone of the square lattice. In this region, the normal-state dispersion around the chemical potential in both quasicrystals -- without RSOC or PZMF (not shown), with RSOC (Fig.~\ref{fig:Dispersion}(b1,c1)), or with RSOC+PZMF (Fig.~\ref{fig:Dispersion}(b2,c2)) -- is similar to that in the square lattice. Clearly seen in Fig.~\ref{fig:Dispersion}(a3,b3,c3) are the chiral edge states in the TSC phase with $B=1$ in all three systems.

\begin{figure*}
  \Large{(a)\qquad\qquad\qquad\qquad \qquad(b)\qquad\qquad\qquad\qquad (c)\qquad\qquad\qquad\qquad\qquad\qquad}\\
  \includegraphics[width=0.6\columnwidth]{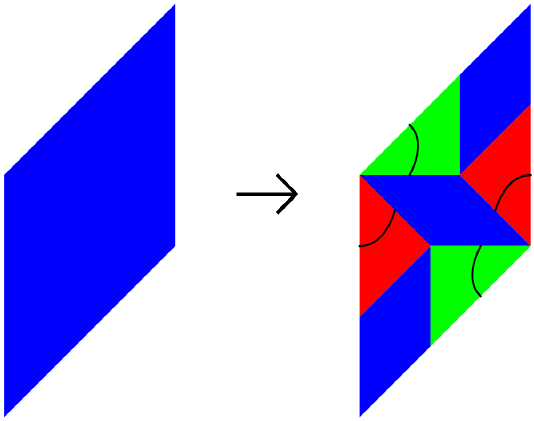}
  \includegraphics[width=0.55\columnwidth]{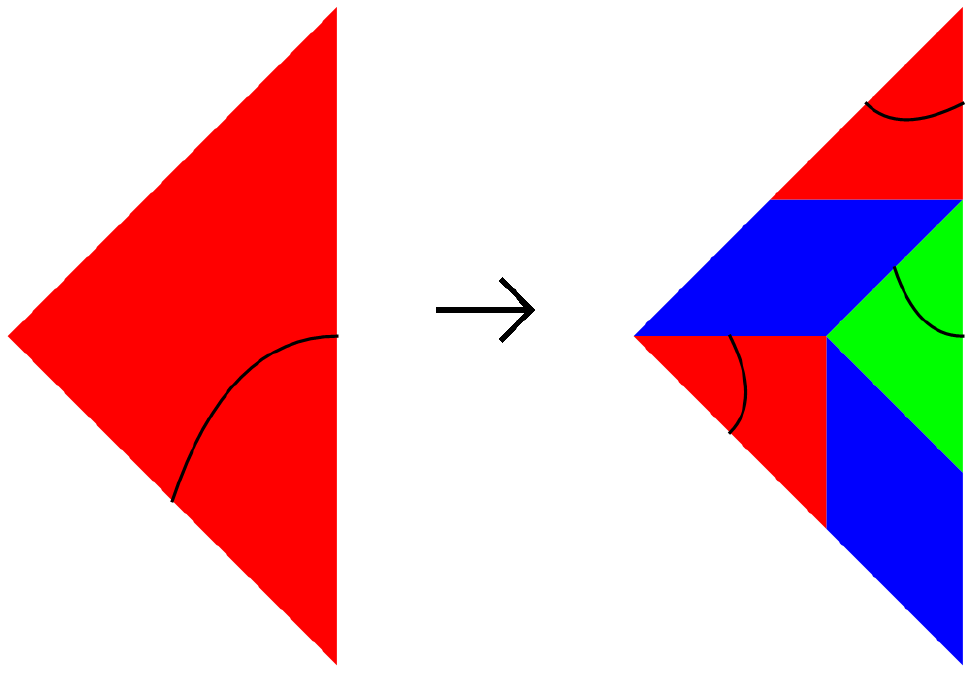}
  \includegraphics[width=0.6\columnwidth]{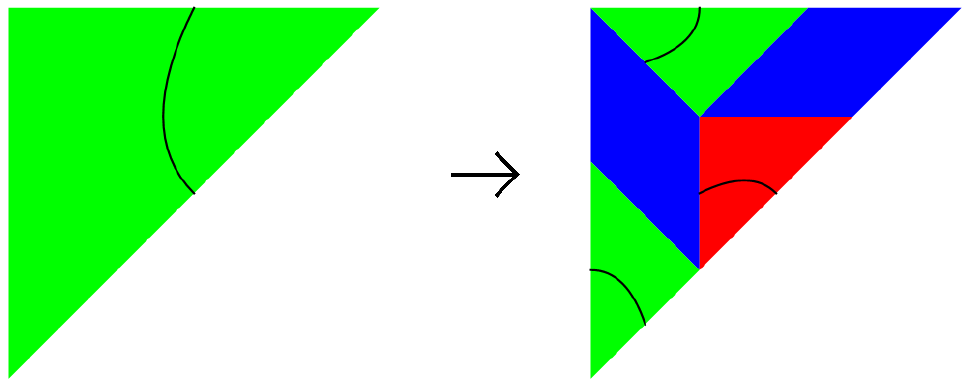}
  \caption[Deflation rules]{Deflation rules for a (a) 45$^\circ$ rhombus and (b), (c) two isosceles right triangles with the (b) left and (c) right corner marked. }
  \label{fig:Deflation}
\end{figure*}

Thus, the topological phase transitions 
as predicted by Eq.~(\ref{murange})
can be understood as due to the fact that the normal-state dispersion is similar in all three systems. One might then ask if the topological phase studied in this work, which does not rely on any crystal symmetry, can occur in arbitrary random structure. Or is there anything special about quasicrystals, and in particular, is bipartite structure key to the existence of TSC in quasicrystals? To address the last question, we have performed additional calculation by making all three systems non-bipartite; by including the next-nearest-neighbor links and the shortest links, respectively, in a square lattice and quasicrystals. The classification of different topological phases for different ranges of $\mu$ \cite{PhysRevB.82.134521} is still possible for a square lattice, even though modified by next-nearest-neighbor hopping.
We have also confirmed that making the Penrose and AB quasicrystals non-bipartite by including the shortest links does not alter the results presented above. Namely, not only the TSC phase with $B=1$ that supports Majorana zero modes, but also the topological phase transitions analogous to those in a square lattice occur in both quasicrystals made non-bipartite with the shortest links.

        \subsection{Effects of randomness} \label{subsec:random}

To gain further insight, we have searched for the TSC phase in all three systems while either introducing random additional links or removing random nearest-neighbor links, so as to gradually approach the extreme limit of arbitrary random structure. In all cases that we have examined
for the same parameters as in Eq.~(\ref{parameters}),
the TSC phase for a range of $\mu$ roughly given by Eq.~(\ref{murange}) persists up to some degree of randomness; however, $\mu$ quickly goes out of this range as randomness increases and different sets of parameters --typically larger $\Delta$-- are required altogether for TSC states, if any, to exist (in different kinetic-energy ranges). 
For example, with longer links added randomly in the Penrose (9349 vertices) and AB (8119 vertices) quasicrystals (diagonal links in wider rhombuses and squares, respectively, in the former and latter),
the range of $\mu$ for which TSC occurs as in Fig.~\ref{fig:evbs} is significantly narrowed and shifted more than halfway up in energy compared to Eq.~(\ref{murange}), when the number of added links is 10\% of the total number of links in both quasicrystals. The changes are more drastic in the AB quasicrystal, in which TSC is mostly gone for $\mu$ given in Eq.~(\ref{murange}) with 10\% added links.
Regardless of energy range, 
there is no such criterion as Eq.~(\ref{murange}) that can predict the occurrence of TSC in partially randomized systems beyond a certain degree of randomness.
We have not found any TSC state in completely random structure.

While quasicrystals lack translational symmetry, we believe that their long-range quasiperiodic order and peculiar rotational symmetry associated with higher-dimensional space group allow the existence of TSC in both quasicrystals. The vertices of the Penrose and AB quasicrystalline lattice can be obtained by projection of a set of five-dimensional and four-dimensional hypercubic lattice points, respectively, onto a two-dimensional plane \cite{PhysRevB.39.10519,Duneau_1989}. It has recently been suggested that the localized states at zero kinetic energy in the Penrose quasicrystal may be protected, so to speak, by topology of its five-dimensional parent \cite{PhysRevB.102.064210}. The Penrose and AB quasicrystals share the general properties of quasiperiodicity and higher-dimensional symmetry. It is intriguing to ask whether or not such general properties can result in some kind of universal or common features in the TSC states among different quasicrystals. At the same time, the Penrose and AB quasicrystals have local pentagonal and octagonal rotational symmetry, respectively. It is interesting to examine their respective fractal structure 
in the perpendicular space, as has recently been done for magnetically ordered states in both quasicrystals \cite{PhysRevB.96.214402,PhysRevB.102.115125}. 
These questions are left for future studies.

	\section{Conclusion} \label{sec:conclusion}
        We have extended the 2D TSC model with broken time-reversal symmetry to 2D quasicrystals. We have shown that despite the aperiodicity, a nontrivial topological phase can be realized in Penrose and Ammann-Beenker quasicrystals at low filling, where the Bott index $B$ is nonzero. By assuming a uniform order parameter $\Delta$, we have observed that topological phase transitions to/from the TSC phase with $B=1$ are governed by Eq.~(\ref{murange}) 
in both quasicrystals, just as for systems with translational symmetry. 
When $B=1$, both 
quasicrystals host chiral surface bound states. 
        Furthermore, by introducing a vortex at the center of the system in the $B=1$ phase, we have found two 
Majorana zero modes, one along the surfaces and the other around the vortex center, irrespective of the underlying crystal structure. In contrast to the CdGM states, the energy of the Majorana vortex bound state 
remains approximately zero regardless of $\mu$ 
and the vortex position. 
        Our results indicate that a new setup of heterostructure using quasicrystals as in Fig.~\ref{fig:crystall}(c) is possible for realizing 
2D TSC.

\begin{figure*}
  \Large{(a)\qquad\qquad\qquad\qquad\qquad (b)\qquad\qquad\qquad\qquad\qquad (c)\qquad\qquad\qquad\qquad\qquad \qquad\qquad}\\
  \includegraphics[width=1.8\columnwidth]{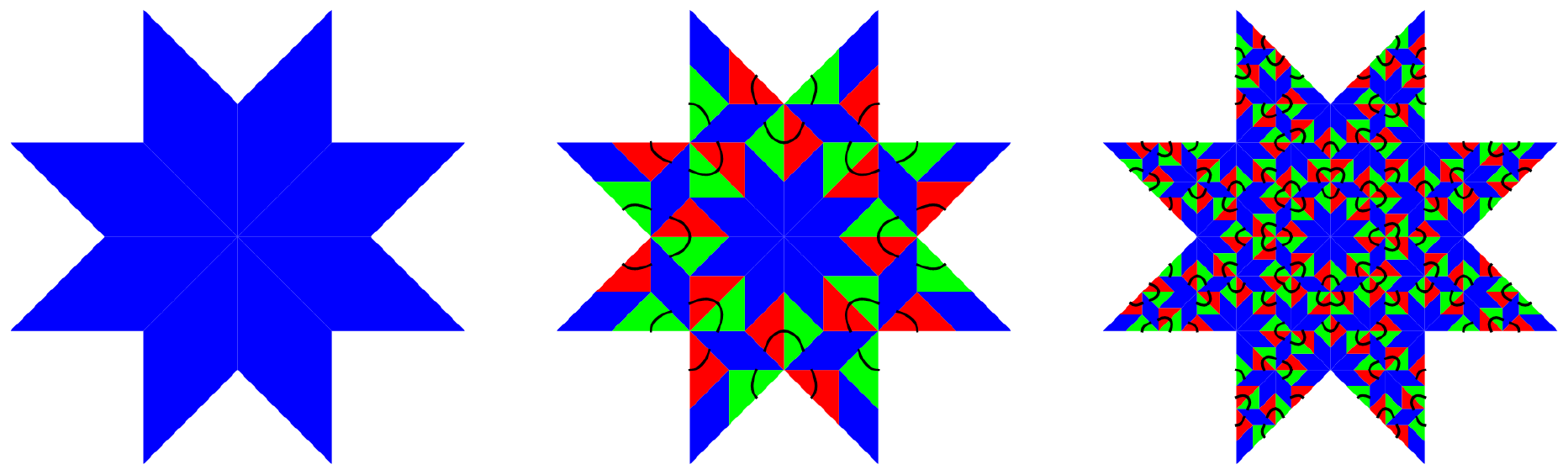}
  \caption[applying deflation rules]{Ammann-Beenker tiling produced by starting from an (a) eightfold-symmetric seed and applying the deflation rules (b) once and (c) twice.}
  \label{fig:inflatedlattice}
\end{figure*}

        Finally, 
it is interesting to explore possible fractal structure in 2D quasicrystals due to their inherent self-similarity, for example, in topological phase diagrams. In order to study such fractal structure, however, the local environment of each vertex in a quasicrystal needs to be taken into account by solving for the order parameter self-consistently. We have performed some preliminary self-consistent calculation of the superconducting order parameter. As found for regular $s$-wave superconductivity \cite{PhysRevB.95.024509,PhysRevB.100.014510}, the order parameter is not uniform in a quasicrystal when solved self-consistently. 
However, spatial fluctuations in the order parameter do not fundamentally alter the conclusions of the current work. By solving for the order parameter self-consistently, we have found the TSC phase with the Bott index $B=1$, where the Majorana zero mode appears along the surface boundary or at the vortex center. Detailed self-consistent studies of TSC in quasicrystals will be presented in a future publication.

\section{Acknowledgments}

        This work was supported by the Japan Society for the Promotion of Science, KAKENHI (Grant No. JP19H05821), and by the Natural Sciences and Engineering Research Council of Canada. K.T. is grateful to the Tokyo University of Science for hospitality, where part of the research was performed. We thank the anonymous referees for constructive criticisms, which have helped improve the manuscript.

\appendix*

\section{Ammann-Beenker approximants}

        We here present our approach to generating the Ammann-Beenker (AB)  \cite{grunbaum1987tilings,Ammann1992,beenker1982algebraic} quasicrystal and its approximants. 
        The Bott index needs to be calculated in an approximant so that the periodic boundary condition (PBC) can be applied.
        With the conventional method of creating approximants by means of projection from a higher-dimensional hypercubic lattice \cite{PhysRevB.39.10519,Duneau_1989}, the number of vertices $N$ in an AB approximant takes on values $N=7, 41, 239, 1393, 8119, 47321, 275807\dots$. Thus, the jump from one possible value of $N$ to the next increases significantly as the system size increases. While an approximant with $N=1393$ is too small to calculate the Bott index for, the next available size $N=8119$ is numerically much more costly to produce.

\begin{figure*}
\Large{(a)\qquad\qquad\qquad\qquad\qquad\qquad\qquad (b)\qquad\qquad\qquad\qquad\qquad\qquad\qquad\qquad\qquad}\\
	\includegraphics[width=0.8\columnwidth]{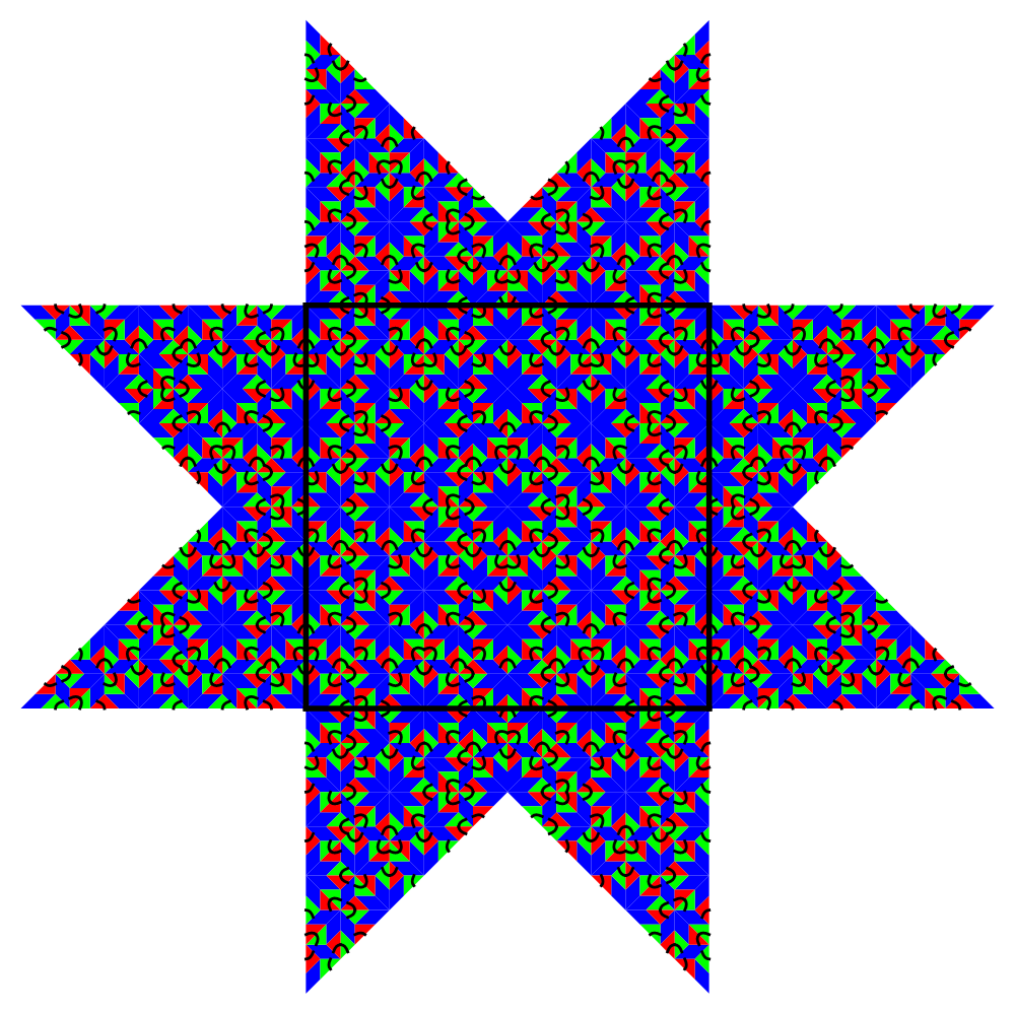}
	\includegraphics[width=0.8\columnwidth]{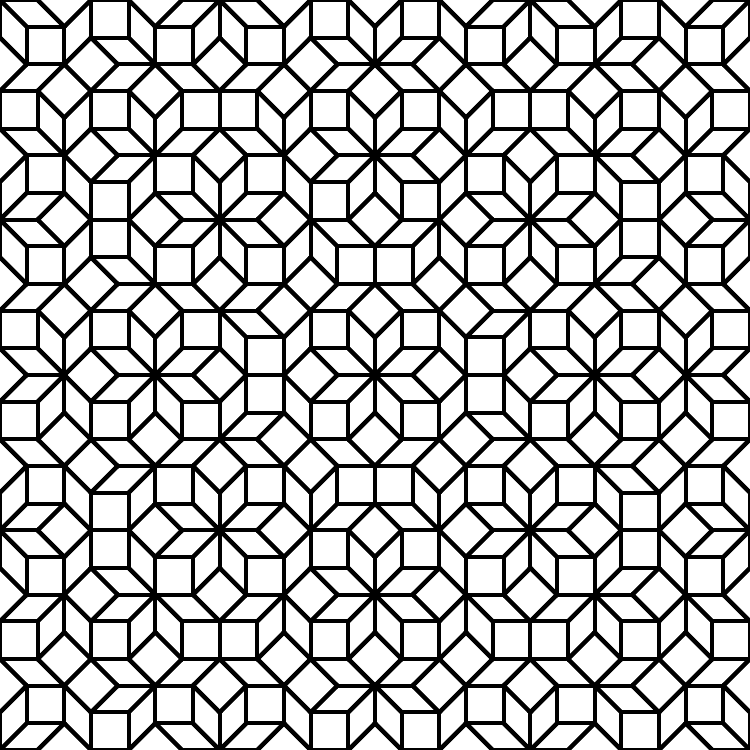}
	\caption[Ammann-Beenker tiling]{(a) Ammann-Beenker tiling after application of the deflation rules to the tiles in Fig.~\ref{fig:inflatedlattice}(c) and (b) the Ammann-Beenker quasicrystal.}
	\label{fig:AmmannBeenker}
\end{figure*}

        For this reason, we use the deflation/inflation rules \cite{Penrose} to produce the AB tiling and identify a large enough square portion enclosed by two sets of parallel Ammann lines \cite{PhysRevB.34.596,PhysRevB.39.10519}. This method allows smaller jumps in possible values of $N$ and hence various intermediate sizes not available with the projection method. We use $N=2786$ for calculation of the Bott index.

        The AB tiling can be constructed in terms of a 45$^\circ$ rhombus as in Fig.~\ref{fig:Deflation}(a) and two isosceles right triangles, represented by red and green triangles with a specific corner marked in Fig.~\ref{fig:Deflation}(b) and (c), respectively. 
One can create the AB tiling by applying the deflation rules to a rhombus and two triangles as shown in Fig.~\ref{fig:Deflation} repeatedly. This is illustrated in Fig.~\ref{fig:inflatedlattice}, where the rules are applied first to each of the eight rhombuses making up the eightfold-symmetric seed tile in panel (a), then to each of the rhombuses and triangles in panel (b), which results in panel (c). Applying the rules once more leads to Fig.~\ref{fig:AmmannBeenker}(a). Thus, more (smaller) tiles are created and the number of vertices increases at each step. 
Alternatively, one can ``inflate'' each side by a factor ($1+\sqrt{2}$) at the same time as the rules are applied, keeping the size of individual tiles the same as the system becomes larger.

\begin{figure*}
\Large{(a)\qquad\qquad\qquad\qquad\qquad\qquad\qquad (b)\qquad\qquad\qquad\qquad\qquad\qquad\qquad\qquad\qquad}\\
	\includegraphics[width=0.9\columnwidth]{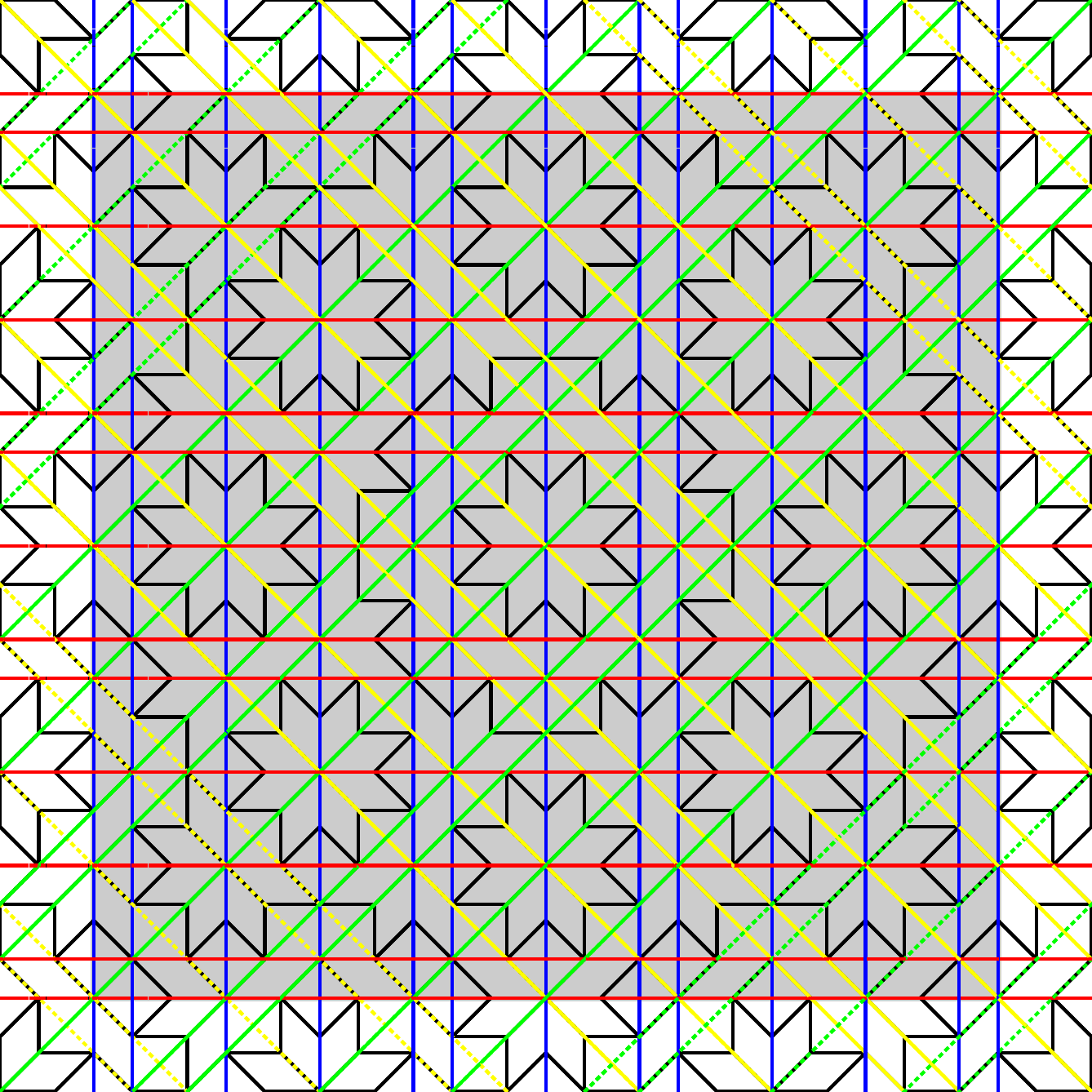}
	\includegraphics[width=0.9\columnwidth]{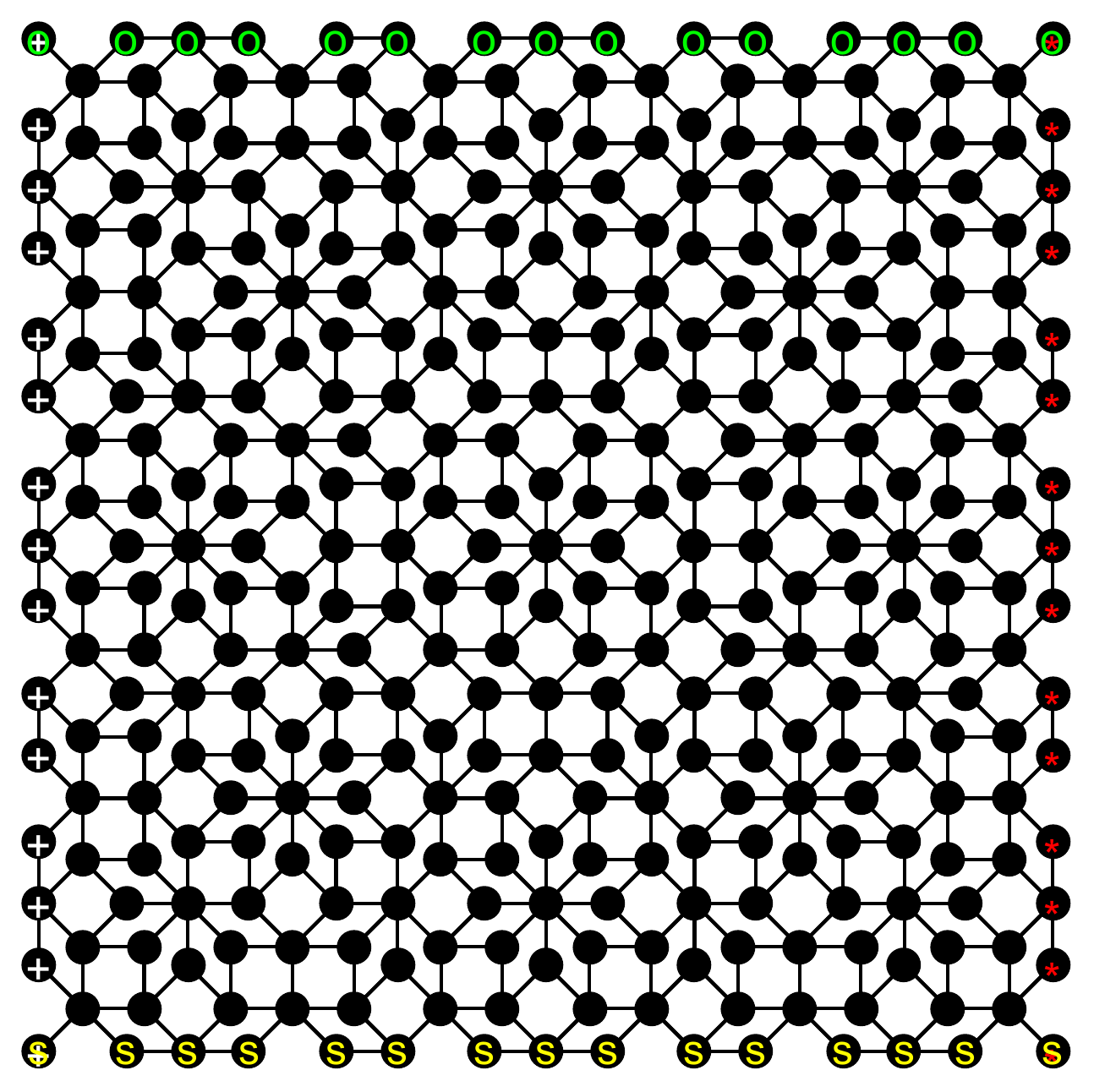}
	\caption[approximant]{(a) Distinct Ammann lines in the Ammann-Beenker quasicrystal are shown in different colors. The gray area is surrounded by two sets of parallel Ammann lines, which have the same vertex configuration along them. (b) An Ammann-Beenker approximant constructed by selecting vertices in the gray region in (a).
	}
	\label{fig:Approximant}
\end{figure*}

We can construct the conventional AB tiling by merging adjacent red and green triangles in Fig.~\ref{fig:AmmannBeenker}(a) into a square tile, so that the building blocks are the square and 45$^\circ$ rhombus \cite{Ammann1992,Bedaride2013,PhysRevLett.59.1010}. The resulting AB quasicrystal surrounded by thick lines in Fig.~\ref{fig:AmmannBeenker}(a) is shown in Fig.~\ref{fig:AmmannBeenker}(b), where the vertices are the quasicrystalline lattice sites and links connecting vertices are all of the same length. 

To calculate the Bott index, we apply PBC to an approximant of the AB quasicrystal. We identify a pair of parallel Ammann lines 
with the same configuration of vertices along them 
and another such pair in the perpendicular direction, to produce an approximant enclosed by these four sets of line segments [the gray region in Fig.~\ref{fig:Approximant}(a)]. Shown in Fig.~\ref{fig:Approximant}(a) are the
Ammann lines as four sets of parallel lines (colored with red, green, blue, and yellow) passing through the diagonal of all square tiles \cite{CAI20192213}.
Figure~\ref{fig:Approximant}(b) illustrates application of PBC in such an approximant, where a vertex with a white plus (green ``O'') is connected to the corresponding vertex with a red star (yellow ``S'') on the other side.

	\bibliography{myreference4}
	
\end{document}